\documentclass[prb]{revtex4}

\usepackage{graphicx}
\usepackage{dcolumn}
\usepackage{bm}

\begin{document}

\title{Influence and Inference in Bell's Theorem}

\author{G. Potvin}

\affiliation{Defence R\&D Canada - Valcartier, Val-B\'{e}lair, Qu\'{e}bec, Canada, G3J 1X5%
\\
}%

\date{\today}

\begin{abstract}
Jaynes' criticism of Bell's theorem is examined.
It is found that Bell's reasoning leading to his theorem is sound when properly interpreted.
The nature of physical influence and logical inference in quantum mechanics, and its relationship with the notions of realism and locality, is discussed.
\end{abstract}

                              
\maketitle

\section{Introduction}

Bell's famous inequalities are derived from the probabilistic reasoning in Bell's theorem \cite{Bell87}, which is based on Bohm's version of the EPR experiment \cite{Bohm51} (denoted as EPR-B).
That reasoning was criticized by Jaynes \cite{Jaynes89} on the grounds that 1) Bell mistakenly equates logical inference with causal influence and that 2) Bell's theorem does not rule out all local hidden variable theories.
Jaynes' views still have some influence today \cite{Kracklauer02,KracklauerKracklauer02}. 
Whether there exists local hidden variable theories that violate Bell's inequalities is a matter of ongoing debate and will not be addressed here.
Rather, we will argue that Jaynes' first criticism proceeds from a misunderstanding of Bell's reasoning.
The next section will review the essential facts concerning the EPR-B experiment.
In section \ref{S_BM} we examine Bell's local realistic model.
Jaynes' views on probability in physics, his objection and responses to it are examined in section \ref{S_IaI}.
We discuss the link between influence and inference with realism and locality in quantum mechanics in section \ref{S_Disc} and we conclude in section \ref{S_Concl}.

\section{The EPR-B Experiment}
\label{S_EPRB}

We begin by recalling the setup of the EPR-Bohm experiment.
Two spin-1/2 particles depart in opposite directions in a singlet state.
We assume that particle 1 moves to the left a distance $L$, where a Stern-Gerlach apparatus (labeled A) measures its spin along the  $\phi_{a}$ direction, where $\phi_{a}$ is the angle of the orientation of the Stern-Gerlach instrument with respect to the ``up" direction.
Similarly, particle 2 moves the right a distance $L$ to another Stern-Gerlach apparatus (B) with an orientation $\phi_{b}$.
The measurement events at A and B are spacelike separated, and we assume that the measurement at A occurs prior to the measurement at B in the reference frame of the experiment.
Each Stern-Gerlach apparatus yields either a spin-up ($u_{a}$, $u_{b}$) or spin-down ($d_{a}$, $d_{b}$) result (with respect to the orientation of each apparatus).\\

We adopt the convention that $P(u_{a},u_{b}|\phi_{a},\phi_{b})$ is the joint conditional probability that both Stern-Gerlach instruments register an ``up" result given the instrument orientations, $\phi_{a}$ and $\phi_{b}$.
The joint conditional probabilities for this experiment are 
\begin{equation}
P(u_{a},u_{b}|\phi_{a},\phi_{b}) = P(d_{a},d_{b}|\phi_{a},\phi_{b})= \frac{1}{2} \sin^{2}(\theta_{ab}/2)
\label{E_Quu}
\end{equation}

\begin{equation}
P(u_{a},d_{b}|\phi_{a},\phi_{b}) = P(d_{a},u_{b}|\phi_{a},\phi_{b})= \frac{1}{2} \cos^{2}(\theta_{ab}/2)
\label{E_Qud}
\end{equation}

where $\theta_{ab} = \phi_{a} - \phi_{b}$.
Because measurement A occurs before measurement B, it is not possible to sort the results of A according to the results at B (at least not yet).
We must therefore take the marginal distribution at A.
The probability for an ``up" result is $P(u_{a}|\phi_{a}) = P(u_{a},u_{b}|\phi_{a},\phi_{b}) + P(u_{a},d_{b}|\phi_{a},\phi_{b}) = 1/2$, and likewise $P(d_{a}|\phi_{a}) = 1/2$.
For the B measurement, we cannot (yet) sort the results according the results at A because the measurement events are spacelike separated and so no information about what happened at A can yet reach B.
The marginal distribution at B is $P(u_{b}|\phi_{b}) = P(d_{b}|\phi_{b}) = 1/2$.
The first thing to notice is that the marginal distribution at A does not depend on the instrument orientation at B, or the results at B, nor even on the instrument orientation at A.
The same can be said about the marginal distribution at B.
It is therefore not possible to send information faster than light using this setup.
Secondly, once we have compiled the results from both A and B, we can see a correlation between them since, clearly, $P(u_{a},u_{b}|\phi_{a},\phi_{b}) \neq P(u_{a}|\phi_{a})P(u_{b}|\phi_{b})$ and the same applies for the other joint conditional probabilities.

\section{Bell's Model}
\label{S_BM}

In this section, we will examine Bell's local realistic model, but with some alterations to promote clarity and simplicity.
The hidden variable(s), that is the variables with well-defined but unknown values that the particles are said to possess in addition to the statevector and independently of any observation, are designated by Bell with the general symbol $\lambda$, which can represent any number of parameters (it can even represent a field).
We will be more specific and assume that particle 1 has a well-defined but unknown spin vector, which is in the plane perpendicular to its motion and forms an angle $\phi_{1}$ with respect to the ``up" direction.
We assume that the spin angle of particle 1 is initially uniformally distributed between 0 and $2\pi$, and that, because of the singlet state, particle 2 has a spin angle $\phi_{2} = \phi_{1} - \pi$.
Finally, we assume that the spin angles do not change as the particles travel towards the Stern-Gerlach instruments.\\

The measurements at A and B must be understood as transformations of the hidden variables.
In other words, the measurement at A is really a \emph{stochastic process} which takes the input variables ($\phi_{a}$, $\phi_{1}$) and changes the particle spin angle into $\phi_{1} = \phi_{a}$ (corresponding to an up result) with a certain probability, or into $\phi_{1} = \phi_{a} - \pi$ (a down result) with the complimentary probability.
Bell used a deterministic measurement process, but we will assume for simplicity that the measurement produces some kind of random disturbance on the system being measured, without going into any kind of detail.
We do require that our model reproduces quantum mechanical probabilities for a single particle.
The measurement probabilities must therefore be same as though we had two Stern-Gerlach instruments in succession, the first one preparing the particle into a well defined spin state and the second one oriented at some angle with respect to the first.
We require therefore,
\begin{equation}
P(u_{a}|\phi_{a},\phi_{1}) = \cos^{2}(\theta_{a1}/2)
\label{E_MeasPro}
\end{equation}
where $\theta = \phi_{a}-\phi_{1}$.
Although Eq. (\ref{E_MeasPro}) has the form of a conditional probability, it is important to emphasize that it represents a dynamical process, i.e. the action of the Stern-Gerlach apparatus on the particle spin angle.
And since the interaction is assumed to be local, Eq. (\ref{E_MeasPro}) cannot depend on anything relating to B.\\

We are now ready to write a general joint probability distribution.
Based on the preceeding description of the stochastic process we use to model the EPR-B experiment, it is clear that,
\begin{equation}
P(u_{a},u_{b},\phi_{1},\phi_{2}|\phi_{a},\phi_{b}) = P(u_{a}|\phi_{a},\phi_{1}) P(u_{b}|\phi_{b},\phi_{2}) \rho(\phi_{1},\phi_{2})
\label{E_BellJ}
\end{equation}
where the joint probability density of the spin angles is $\rho(\phi_{1},\phi_{2}) = (2 \pi)^{-1} \delta (\phi_{1}-\phi_{2}-\pi)$.
Since we don't know the initial conditions of the spin angles, we integrate them out to obtain,
\begin{equation}
P(u_{a},u_{b}|\phi_{a},\phi_{b}) = P(d_{a},d_{b}|\phi_{a},\phi_{b})= \frac{1}{8}+\frac{1}{4}\sin^{2}(\theta_{ab}/2)
\end{equation}
\begin{equation}
P(u_{a},d_{b}|\phi_{a},\phi_{b}) = P(d_{a},u_{b}|\phi_{a},\phi_{b})= \frac{1}{8}+\frac{1}{4}\cos^{2}(\theta_{ab}/2)
\end{equation}
which is not the same as in Eqs. (\ref{E_Quu}) and (\ref{E_Qud}).
This shows that local realistic models of the kind described here do not reproduce quantum mechanical statistics for multiparticle states even if they reproduce them for single particle states.

\section{Influence and Inference}
\label{S_IaI}

\subsection{Logic, probability and relative frequencies}
\label{Sb_LPRF}

Here we briefly review some of Jaynes' ideas on the nature of probability that are relevant to the EPR experiment.
Jaynes advocated the view that probabilities are essentially our state of knowledge (or inference) about an outside, realistic and presumably deterministic world.
This state of knowledge is a kind of `extended logic' using Bayesian inference and maximum entropy (ME) constrained by all available information (the word `entropy' is to be understood in the information theoretic sense introduced by Shannon \cite{Shannon48}).
For Jaynes, this view began in the 1950's when he applied it to classical \cite{Jaynes57a} and quantum \cite{Jaynes57b} statistical mechanics, and culminating in his posthumously published book \cite{Jaynes03} in 2003.
Over this period, Jaynes' views on probabilities in physics underwent sublte and sometimes unexplained changes (see Guttmann \cite{Guttmann99} for an overview and critique of Jaynes' theory).
For our purposes, we will take Jaynes' book as his final word on these matters.
Bayesian inference and ME, therefore, constitute a form of logic extended to deal with uncertainty, and just as in conventional logic, it is necessarily true.
In other words, after we maximize the entropy of a distribution subject to all the \emph{relevant} information we have about the problem (and there is no clear criterion for relevancy), the statistics we then collect concerning the variable in question ought to conform to the ME distribution, but if it does not, then for Jaynes that can only mean that we have not taken into account all the relevant information and should attempt to discover the missing information.
This extended logic can therefore be used to discover previously hidden properties, laws and/or entities.
Furthermore, these logical (or  inferential) probability assignments can apply to a single unreproducible event and do not necessarily depend on any notion of relative frequencies or ergodicity.\\

Jaynes therefore draws a distinction between logical probabilities (which he just called probabilities) and `random variables' that exhibit relative frequencies upon the repetition of a `random' experiment (i.e. `physical' probabilities which, for Jaynes, do not exist).
Although Jaynes developed his theory to deal mainly with statistical mechanics, it is clear he believed it applies equally well to quantum mechanics \cite{Jaynes03}.
In what follows, we will not address the validity of the ME principle in quantum mechanics (and so leave aside claims that minimizing the Fisher information gives better results \cite{Frieden90,Luo02}), and concentrate only on the nature of probabilities (physical or inferential) in quantum mechanics.

\subsection{Jaynes and the EPR experiment}
\label{Sb_JEPR}

Jaynes' main contention was that Bell's factorization (\ref{E_BellJ}) does not follow from the rules of probability theory, according to which the fundamentally correct factorization should be,
\begin{equation}
P(u_{a},u_{b},\phi_{1},\phi_{2}|\phi_{a},\phi_{b}) = P(u_{a}|u_{b},\phi_{a},\phi_{b},\phi_{1},\phi_{2}) P(u_{b}|\phi_{a},\phi_{b},\phi_{1},\phi_{2}) \rho(\phi_{1},\phi_{2})
\label{E_Jaynes1}
\end{equation}
or,
\begin{equation}
P(u_{a},u_{b},\phi_{1},\phi_{2}|\phi_{a},\phi_{b}) = P(u_{a}|\phi_{a},\phi_{b},\phi_{1},\phi_{2}) P(u_{b}|u_{a},\phi_{a},\phi_{b},\phi_{1},\phi_{2}) \rho(\phi_{1},\phi_{2}).
\label{E_Jaynes2}
\end{equation}
Interestingly, the key to the disagrement between Bell and Jaynes was given by Jaynes himself \cite{Jaynes89}.
In order to obtain a rational picture of the world, Jaynes emphasized the need to distinguish between reality (as expressed by the laws of physics describing physical causation at the level of ontology) and our knowledge of reality (where probability theory describes human inference at the level of epistemology).\\

It is clear from the previous section, however, that Bell's factorization (\ref{E_BellJ}) was built ``from the ground up".
That is, we first considered a realistic, local stochastic process with random initial conditions as a description of what is going on in the EPR-B experiment at the ontological level, and from this we deduced Eq. (\ref{E_BellJ}).
It therefore represents (a hypothetical) reality.

\subsubsection{Dice and randomization}
\label{Sbb_DAR}

Jaynes' factorizations, Eq. (\ref{E_Jaynes1}) and (\ref{E_Jaynes2}), on the other hand, represent our knowledge of reality.
They are based solely on probability theory and make no physical assumptions.
They are the most general inference we can make.
But since Bell was trying to evaluate the statistical consequences of a specific physical model, they are too general.
As an example, let us consider three dice (which may or may not be fair): a red die (with outcome labelled $n_{r}$), a green die ($n_{g}$) and a blue die ($n_{b}$).
We shake the dice in a cup and throw them on a table.
Probability theory tells us that the probability for the result of a throw, $P(n_{r},n_{g},n_{b})$, can be factorized as,
\begin{equation}
P(n_{r},n_{g},n_{b}) = P(n_{r}|n_{g},n_{b})P(n_{g}|n_{b})P(n_{b}).
\label{E_diceG}
\end{equation}
However, we know from the physical process of the throw (the dice were shaken for a long time, they rolled on the table over a considerable distance, etc) that we can factorize the joint probability as,
\begin{equation}
P(n_{r},n_{g},n_{b}) = P(n_{r})P(n_{g})P(n_{b}).
\label{E_diceR}
\end{equation}
Equation (\ref{E_diceR}) is a special case of Eq. (\ref{E_diceG}) and does not contradict it.
It is a special case because it incorporates physical information that Eq. (\ref{E_diceG}) does not contain.
The same thing can be said about Bell's factorization (\ref{E_BellJ}) with respect to Jaynes' factorization, (\ref{E_Jaynes1}) and (\ref{E_Jaynes2}).\\

It is here that Jaynes would take us to task.
We assumed that the `randomization' of the dice is a physical process in of itself.
In fact, the shaking and tossing is a physical process obeying the laws of classical physics.
If we denote the exact details of the tossing process as $T$, and our knowledge about that process as $\rho(T|I)$, where $I$ is the information we have about $T$, then our knowledge of the outcome of the toss is,
\begin{equation}
P(n_{r},n_{g},n_{b},T|I) = P(n_{r}|n_{g},n_{b},T)P(n_{g}|n_{b},T)P(n_{b}|T)\rho(T|I).
\label{E_diceRT}
\end{equation}
We then integrate Eq. (\ref{E_diceRT}) over $T$ to obtain a marginal distribution that, owing to the intricate relationship between $T$ and the outcome and our very incomplete knowledge of $T$, looks very much like Eq. (\ref{E_diceR}).
But Eq. (\ref{E_diceR}) is here a logical probability.
Should we actually throw the dice a large number of times, and should there be an unknown systematic influence acting on the tossing process, the relative frequencies may look very different from Eq. (\ref{E_diceR}).
In such a case, we must revise our knowledge of $T$ until we discover the systematic influence.
This example also illustrates Jaynes' view of the origin of randomness in quantum mechanics, \cite{Jaynes03} i.e. as a consequence of our incomplete knowledge of hidden causes $C$ which determine the outcomes of quantum measurements.

\subsubsection{Classical urns and quantum gloves}
\label{Sbb_CUAQG}

Jaynes attempts to show that conditional probabilities such as $P(u_{a}|u_{b},\phi_{a},\phi_{b},\phi_{1},\phi_{2})$ or $P(u_{b}|u_{a},\phi_{a},\phi_{b},\phi_{1},\phi_{2})$ do not imply action-at-a-distance by invoking Bernoulli's urn.
Consider an urn containing $N$ identical balls except that $M$ of them are red and the remaining $N-M$ are white.
The balls are randomly drawn out one at a time and not replaced.
The probability that the first ball is red is $M/N$.
If the first ball is red, then the conditional probability that the second ball is red is $(M-1)/(N-1)$.
This conditional probability represents the causal influence the first ball has on the second ball.
But suppose that we discarded the first ball without looking at it.
Not knowing whether the first ball was red or white, we can only say that the probability that the second ball is red is $M/N$.
If the second ball is red, then the conditional probability that the first ball was red is $(M-1)/(N-1)$.
This second conditional probability represents our inference regarding a past event and not a causal influence, although it is identical to the previous conditional probability.\\

As instructive as the Bernoulli's urn example is, it does not fully capture the quantum situation.
The most important difference is that in a quantum experiment, the act of measurement itself can transform the object being measured.
This would be like a red ball spontaneously becoming a white ball with a probability that depends on how it is observed (the colour of the ball is therefore a \emph{contextual} property relative to the experimental circumstances of observation).
We can pursue the analogy with macroscopic objects using Lindley's \cite{Lindley96} glove example.
Consider first a pair of ``classical" gloves.
One glove (either the left or right handed glove) is placed in box A, and the other glove is placed in box B.
We will modify Lindley's example by assuming that the gloves are fastened to the bottom of the boxes, and that the boxes are equipped with several windows closed by a system of shutters.
The windows allow one to see the gloves at different angles, and the shutter mechanism is such that only one shutter can be opened at a time.
The two boxes are transported far away from each other.
The observer carrying box A opens a shutter and sees a left-handed glove.
He therefore instantly knows (or infers) that box B contains a right-handed glove, regardless of how far away it may be or if one of the shutters on box B has been opened.
This example does not call into question realism or locality due to the nature of observation in the classical world.
The results of observation are independent of how the observation is carried out.
If the shutter in box A is closed and another one opened, the observer still sees a left-handed glove with a probability of 1.
It is this fact that allows us to attribute an independent reality to the gloves and their properties.
Since the gloves are real and are not affected by the observations made on them, the results of those observations can only refer to the knowledge the observers have on this independent reality.
There is therefore no faster-than-light causal influence in this case.\\

Now we consider the case of ``quantum" gloves.
First, we have to contend with the fact that observation can change what is being looked at.
If we open the shutter on the top of box A and see a left-handed glove, then we close that shutter and open the shutter on the right side of box A, there is now a 50 percent probability of seeing a right-handed glove.
Second, we must consider the fact that the two gloves form a pair.
In a fashion-conscious variant of quantum terminology, we would say that both gloves are `entangled' in a `pair state'.
The `pairness' is, in a sense, a property of neither glove individually, but of both gloves together (it is therefore what Philippe Grangier \cite{Grangier03} would call a ``holistic" property).
We are now ready to see how realism, in this context, implies non-locality.
Let us assume that both gloves possess a definite handedness at all times (though we may not know which).
For simplicity, we assume that both observers open the right-sided shutter, the observer at box A performing the operation just prior to the observer at box B.
Now let us suppose that originally the left-handed glove was in box A, but that the act of observing it changed it to a right-handed glove.
Since the observer at box B is observing the glove in the same manner (he is opening the same shutter), he must find a left-handed glove with a probability of 1 on account of the pair state of the gloves.
This means that some sort of causal influence must have travelled faster than light, from box A to box B, to cause the glove in box B to change handedness.
It is the preservation of this pair state that is lacking in Bell's model, where it is perfectly possible for both observers to find left or right-handed gloves.
It is important to note that the changes caused by measurement, both for the gloves and Bell's model, are assumed genuinely and irreducibly random.
However, even if we use hidden causes to account for the randomness we will see that, under certain reasonable assumptions, these causes do not appear to alter the content of Bell's theorem. 

\section{Discussion}
\label{S_Disc}

\subsection{Locality, realism and Bell's theorem}
\label{Sb_LRBT}

The last paragraph of subsection \ref{Sbb_CUAQG} brings us to the crux of the matter.
It is by assuming that the gloves always have a definite handedness, and taking into account the entanglement of the gloves and the disruptive nature of a quantum measurement, that we deduce the \emph{possibility} of a non-local interaction having taken place.
However, all we really know are the results of our measurements.
Perhaps observing the gloves did not change their handedness, in which case our observations merely revealed the pre-existing situation and nothing non-local happened.
In any case, the situation of the gloves \emph{prior} to our observations is, by definition, unobservable.
It is not surprising that realist assumptions lead to the existence of unobservable entities or variables.
Realistic theories attempt to give a complete description of the world at the causal and ontological level.
Since our knowledge of the world is always incomplete and sometimes indirect, it is logical that realistic theories contain entities or variable that were not, or even cannot be, observed.\\

One alternative is to abandon, in a sense, realism to save locality.
Here we must be careful and ask ourselves why do we want to preserve locality if not to preserve the reality of relativistic space-time?
It would be fairer to say that we are abandoning realism with respect to material particles in order to preserve realism with respect to space-time.
One possibility would be to say that the gloves are initially in an indeterminate state, neither left or right-handed.
The act of observation causes a `collapse' of the quantum state and the gloves acquire a definite handedness with a certain probability.
This is the point of view introduced by von Neumann \cite{Neumann55}.
But since the gloves are entangled, the collapse of one implies the collapse of the other, presumably at the same instant (there is no clear criteria for when the collapse occurs).
Therefore, a kind of non-locality still persists.
Recent attempts to replace the collapses with environment-induced decoherence \cite{Zurek91} has been found wanting. \cite{Adler01}\\

Another possibility is adopt the point of view of Neils Bohr in which the quantum system is treated as a `black box'.
Here, we make no assertions about any underlying reality, whether it exists or not and whether it is local or not.
Quantum mechanics is really about laboratory experiments, where the quantum system is first prepared (giving rise to a specific wavefunction).
The experiment ends when the results are obtained (which obey Born's statistical postulate, for a review of this interpretation see Stapp \cite{Stapp72}).
This interpretation says that science is basically about the observations we can make and not about an underlying reality.
To make assumptions about an underlying reality is unwarranted and runs the risk of producing apparent paradoxes.
Here, the wavefunction is not universal but a by-product of the experimental arrangement (i.e. it does not exist outside the context of the preparation-measurement procedure).
Quantum mechanics is therefore local in a purely operational sense by virtue of the impossibility of faster-than-light signalling.
However, this point of view forces us to divide the universe into the quantum system being measured and the classical instruments involved in the experiment.
As Bell once objected \cite{Bell90}, this boundary is not well-defined.
Furthermore, by refusing to say anything about what is happening behind the phenomena, the Copenhagen interpretation avoids paradox by avoiding explanation.\\

The best known non-local realist interpretation of quantum mechanics is undoubtedly Bohmian mechanics, named after Bohm's \cite{Bohm52a, Bohm52b} account of de Broglie's pilot-wave idea (see Holland \cite{Holland93} for an extensive review of Bohmian mechanics, including its treatment of the EPR experiment).
In this theory, the familiar quantum mechanical formalism is supplemented with particle positions, or field configurations for quantum field theory.
In the particle case, the wavefunction is seen as a real field existing in configuration space that guides the particles in a deterministic manner along non-classical trajectories that reproduce all the predictions of standard quantum theory.
Probabilities are introduced by assuming the initial particle configurations are distributed according to Born's rule.
The trajectories are such as to preserve Born's rule for all time.\\

Apart from its great conceptual clarity, Bohmian mechanics also solves the `measurement problem'.
It describes both the macroscopic apparatus and the quantum system under study in the same way, and position measurements simply reveal the pre-existing positions of the particles.
All other types of measurements, such as spin measurements, are really position measurements.
The Stern-Gerlach apparatus causes the wavefunction to branch out into two channels, the wavefunction then guides the particle into one or the other channel and it is the particle's position which is ultimately measured.
Spin is inferred from the position measurement.
Bohmian mechanics being non-local, what happens to the particle at one end of the EPR-B experiment instantly influences what happens to the particle at the other end in such a way as to reproduce the statistics of quantum mechanics (Eqs. (\ref{E_Quu}) and (\ref{E_Qud})).
Although its non-locality implies the existence of a preferred frame, the statistical results of Bohmian mechanics forbides faster-than-light signalling.
There is consequently a peaceful co-existence between Bohmian mechanics and relativity, even though the former violates the spirit of the latter.
Here we see that Bohmian mechanics does not escape our previous conlcusions about realistic theories: conceptual clarity is achieved by introducing undetectable entities (particle trajectories and a preferred frame).

\subsection{Beyond Bell and back to Jaynes}
\label{BBT}

Bell's theorem assumes both realism (in the form of the hidden variables) and locality.
It also depends on other assumptions wich can be relaxed in order to save local realism.
One can, for instance, suppose that the experimenters are not free to choose the settings on their respective instruments.
They are predetermined to choose those settings which, along with the local hidden variables, produces the correct statistics.
Alternatively, we can postulate some sort of backward causation where the measurement on one particle produces a causal influence that travels backward in time to the point of origin of the two particles, and then travels forward in time to the other particle.
We can also suppose that the instruments are not as distant as they appear since space is not really flat but riddled with wormholes \cite{Holland93}.
Finally, we can preserve local realism by using some exotic form of probability theory, such as postulating the validity of negative probabilities \cite{Muckenheim.et.al86} (whatever that means).\\

Bearing this in mind, we revisit Jaynes' logical inference model of quantum probabilities by invoking hidden causes for the preparation procedure ($C_{0}$) and the measurements ($C_{a}$ and $C_{b}$).
We further assume that the preparation procedure is a deterministic process, $\phi_{1}(C_{0})$ and $\phi_{2}(C_{0})$, and so are spin measurements, $s_{a}(\phi_{a},\phi_{1},C_{a})$ and $s_{b}(\phi_{b},\phi_{2},C_{b})$, where $s = [u,d]$.
The probability of obtaining two up results is,
\begin{equation}
P(u_{a},u_{b}|\phi_{a},\phi_{b}) = \int \delta_{k} [u_{a},s_{a}] \delta_{k} [u_{b},s_{b}] \rho(C_{0},C_{a},C_{b}|\phi_{a},\phi_{b},I) {\rm d}C_{0} {\rm d}C_{a} {\rm d}C_{b}
\label{E_Juu}
\end{equation}
where $\delta_{k}[u,s]$ is to be understood as a Kronecker delta function ($\delta_{k}=1$ if $s=u$, zero otherwise).
The probability density $\rho(C_{0},C_{a},C_{b}|\phi_{a},\phi_{b},I)$ represents the logical inferences we can make regarding the causes $C$ based on our knowledge of the settings $\phi_{a}$ and $\phi_{b}$, and on any other relevant information we may have $I$ (which is in no way related, directly or indirectly, to the settings).
This very general probability density can reproduce the quantum mechanical results, but we must question its physical meaning in light of our assumption on the free will of the experimenters.
Logically, there should not be any relationship between the settings and the hidden causes if the experimenters are truly `free'.
To suppose otherwise is to imply backward causation or predetermination.
We must therefore use the probability density $\rho(C_{0},C_{a},C_{b}|I)$ in Eq. (\ref{E_Juu}), which would eliminate any contextuality in this model.\\

The number of possible models one could imagine is quite large (as they depend on the exact nature of the causes and the information we have on them), and so we will not attempt a formal proof.
Rather we will argue that the non-contextual hidden variable model in Sec. \ref{S_BM} depends only on two random variables: $\phi_{1}$, which along with the setting $\phi_{a}$ determined the probability of a spin result at A, and $\phi_{2}$, which along with $\phi_{b}$ did the same at B.
Furthermore, $\phi_{1}$ and $\phi_{2}$ were perfectly anti-correlated.
In our hidden cause model, the spin result at A depends on the random variables $C_{a}$ and $C_{0}$, along with the setting $\phi_{a}$, whereas at B the results depends on the random variables $C_{b}$ and $C_{0}$, along with $\phi_{b}$.
It would seems therefore that the two models are fundamentally the same except that the hidden cause model is more complicated with more random variables that may not be strongly correlated with each other (this is particularly true if we use the ME method to deduce $\rho(C_{0},C_{a},C_{b}|I)$, which tends to suppress correlations).
Without a specific physical model at hand, which is not too contrived or \emph{ad hoc}, we can only conclude that Jaynes' views on probability do not, in general, allow a local realistic account of the EPR-B experiment.

\section{Conclusions}
\label{S_Concl}

The disagreement between Bell and Jaynes may be seen as a conflict between two approaches to doing physics.
Bell used a ``bottom-up" approach, where a realistic model of what is going on behind the phenomena is postulated, its observational consequences are derived and compared with actual observations.
Jaynes prefers a ``top-down" approach, where macroscopic observations come first and it is only on the basis of these observation that one can logically infer the existence of independently existing microscopic entities and their properties.
In classical physics these two approaches are entirey compatible, whereas in quantum physics the nature of measurements makes them difficult to reconcile.
It is this difference in perspectives that may have led Jaynes to overlook the physical content in Bell's theorem.
Jaynes seems to have believed in local realism and that the apperance of non-locality in quantum mechanics was the result of poorly applied inference.
But even when taking into account his views on the origin of probabilities in physics, it seems unlikely that they lead to a local realistic account of the EPR-B experiment.

\section*{Acknowledgments}

The author would like to thank Gwedoline Simon from St-Lawrence College for helpful comments and support.

\bibliography{IIB4}

\begin{thebibliography}{23}
\expandafter\ifx\csname natexlab\endcsname\relax\def\natexlab#1{#1}\fi
\expandafter\ifx\csname bibnamefont\endcsname\relax
  \def\bibnamefont#1{#1}\fi
\expandafter\ifx\csname bibfnamefont\endcsname\relax
  \def\bibfnamefont#1{#1}\fi
\expandafter\ifx\csname citenamefont\endcsname\relax
  \def\citenamefont#1{#1}\fi
\expandafter\ifx\csname url\endcsname\relax
  \def\url#1{\texttt{#1}}\fi
\expandafter\ifx\csname urlprefix\endcsname\relax\def\urlprefix{URL }\fi
\providecommand{\bibinfo}[2]{#2}
\providecommand{\eprint}[2][]{\url{#2}}

\bibitem[{\citenamefont{Bell}(1987)}]{Bell87}
\bibinfo{author}{\bibfnamefont{J.~S.} \bibnamefont{Bell}},
  \emph{\bibinfo{title}{Speakable and Unspeakable in Quantum Mechanics}}
  (\bibinfo{publisher}{Cambridge University Press},
  \bibinfo{address}{Cambridge}, \bibinfo{year}{1987}), \bibinfo{note}{pp.
  14--21}.

\bibitem[{\citenamefont{Bohm}(1951)}]{Bohm51}
\bibinfo{author}{\bibfnamefont{D.}~\bibnamefont{Bohm}},
  \emph{\bibinfo{title}{Quantum Theory}} (\bibinfo{publisher}{Dover},
  \bibinfo{address}{New York}, \bibinfo{year}{1951}).

\bibitem[{\citenamefont{Jaynes}(1989)}]{Jaynes89}
\bibinfo{author}{\bibfnamefont{E.~T.} \bibnamefont{Jaynes}}, in
  \emph{\bibinfo{booktitle}{Maximum Entropy and Bayesian Methods}}, edited by
  \bibinfo{editor}{\bibfnamefont{J.}~\bibnamefont{Skilling}}
  (\bibinfo{publisher}{Kluwer Academic Publishers},
  \bibinfo{address}{Dordrecht-Holland}, \bibinfo{year}{1989}), pp.
  \bibinfo{pages}{1--27}, \bibinfo{note}{also available at the web site
  http://bayes.wustl.edu/}.

\bibitem[{\citenamefont{Kracklauer}()}]{Kracklauer02}
\bibinfo{author}{\bibfnamefont{A.~F.} \bibnamefont{Kracklauer}},
  \eprint{quant-ph/0210121}.

\bibitem[{\citenamefont{Kracklauer and
  Kracklauer}(2002)}]{KracklauerKracklauer02}
\bibinfo{author}{\bibfnamefont{A.~F.} \bibnamefont{Kracklauer}}
  \bibnamefont{and} \bibinfo{author}{\bibfnamefont{N.~A.}
  \bibnamefont{Kracklauer}}, \bibinfo{journal}{Phys. Essays}
  \textbf{\bibinfo{volume}{15}}, \bibinfo{pages}{162} (\bibinfo{year}{2002}).

\bibitem[{\citenamefont{Shannon}(1948)}]{Shannon48}
\bibinfo{author}{\bibfnamefont{C.~E.} \bibnamefont{Shannon}},
  \bibinfo{journal}{Bell System Tech. J.} \textbf{\bibinfo{volume}{27}},
  \bibinfo{pages}{379} (\bibinfo{year}{1948}).

\bibitem[{\citenamefont{Jaynes}(1957a)}]{Jaynes57a}
\bibinfo{author}{\bibfnamefont{E.~T.} \bibnamefont{Jaynes}},
  \bibinfo{journal}{Phys. Rev.} \textbf{\bibinfo{volume}{106}},
  \bibinfo{pages}{620} (\bibinfo{year}{1957a}).

\bibitem[{\citenamefont{Jaynes}(1957b)}]{Jaynes57b}
\bibinfo{author}{\bibfnamefont{E.~T.} \bibnamefont{Jaynes}},
  \bibinfo{journal}{Phys. Rev.} \textbf{\bibinfo{volume}{108}},
  \bibinfo{pages}{171} (\bibinfo{year}{1957b}).

\bibitem[{\citenamefont{Jaynes}(2003)}]{Jaynes03}
\bibinfo{author}{\bibfnamefont{E.~T.} \bibnamefont{Jaynes}},
  \emph{\bibinfo{title}{Probability Theory}} (\bibinfo{publisher}{Cambridge
  University Press}, \bibinfo{address}{Cambridge}, \bibinfo{year}{2003}).

\bibitem[{\citenamefont{Guttmann}(1999)}]{Guttmann99}
\bibinfo{author}{\bibfnamefont{Y.~M.} \bibnamefont{Guttmann}},
  \emph{\bibinfo{title}{The Concept of Probability in Statistical Physics}}
  (\bibinfo{publisher}{Cambridge University Press},
  \bibinfo{address}{Cambridge}, \bibinfo{year}{1999}).

\bibitem[{\citenamefont{Luo}(2002)}]{Luo02}
\bibinfo{author}{\bibfnamefont{S.}~\bibnamefont{Luo}}, \bibinfo{journal}{Found.
  Phys.} \textbf{\bibinfo{volume}{32}}, \bibinfo{pages}{1757}
  (\bibinfo{year}{2002}).

\bibitem[{\citenamefont{Frieden}(1990)}]{Frieden90}
\bibinfo{author}{\bibfnamefont{B.~R.} \bibnamefont{Frieden}},
  \bibinfo{journal}{Phys. Rev. A} \textbf{\bibinfo{volume}{41}},
  \bibinfo{pages}{4265} (\bibinfo{year}{1990}).

\bibitem[{\citenamefont{Lindley}(1996)}]{Lindley96}
\bibinfo{author}{\bibfnamefont{D.}~\bibnamefont{Lindley}},
  \emph{\bibinfo{title}{Where Does the Weirdness Go?}}
  (\bibinfo{publisher}{BasicBooks}, \bibinfo{address}{New York},
  \bibinfo{year}{1996}).

\bibitem[{\citenamefont{Grangier}()}]{Grangier03}
\bibinfo{author}{\bibfnamefont{P.}~\bibnamefont{Grangier}},
  \eprint{quant-ph/0301001}.

\bibitem[{\citenamefont{von Neumann}(1955)}]{Neumann55}
\bibinfo{author}{\bibfnamefont{J.}~\bibnamefont{von Neumann}},
  \emph{\bibinfo{title}{Mathematical Foundations of Quantum Mechanics}}
  (\bibinfo{publisher}{Princeton University Press},
  \bibinfo{address}{Princeton}, \bibinfo{year}{1955}).

\bibitem[{\citenamefont{Zurek}(1991)}]{Zurek91}
\bibinfo{author}{\bibfnamefont{W.~H.} \bibnamefont{Zurek}},
  \bibinfo{journal}{Physics Today} \textbf{\bibinfo{volume}{44}},
  \bibinfo{pages}{36} (\bibinfo{year}{1991}).

\bibitem[{\citenamefont{Adler}()}]{Adler01}
\bibinfo{author}{\bibfnamefont{S.~L.} \bibnamefont{Adler}},
  \eprint{quant-ph/0112095}.

\bibitem[{\citenamefont{Stapp}(1972)}]{Stapp72}
\bibinfo{author}{\bibfnamefont{H.~P.} \bibnamefont{Stapp}},
  \bibinfo{journal}{Am. J. Phys.} \textbf{\bibinfo{volume}{40}},
  \bibinfo{pages}{1098} (\bibinfo{year}{1972}).

\bibitem[{\citenamefont{Bell}(1990)}]{Bell90}
\bibinfo{author}{\bibfnamefont{J.~S.} \bibnamefont{Bell}},
  \bibinfo{journal}{Physics World} \textbf{\bibinfo{volume}{3}},
  \bibinfo{pages}{33} (\bibinfo{year}{1990}).

\bibitem[{\citenamefont{Bohm}(1952a)}]{Bohm52a}
\bibinfo{author}{\bibfnamefont{D.}~\bibnamefont{Bohm}}, \bibinfo{journal}{Phys.
  Rev.} \textbf{\bibinfo{volume}{85}}, \bibinfo{pages}{166}
  (\bibinfo{year}{1952a}).

\bibitem[{\citenamefont{Bohm}(1952b)}]{Bohm52b}
\bibinfo{author}{\bibfnamefont{D.}~\bibnamefont{Bohm}}, \bibinfo{journal}{Phys.
  Rev.} \textbf{\bibinfo{volume}{85}}, \bibinfo{pages}{180}
  (\bibinfo{year}{1952b}).

\bibitem[{\citenamefont{Holland}(1993)}]{Holland93}
\bibinfo{author}{\bibfnamefont{P.~R.} \bibnamefont{Holland}},
  \emph{\bibinfo{title}{The Quantum Theory of Motion}}
  (\bibinfo{publisher}{Cambridge University Press},
  \bibinfo{address}{Cambridge}, \bibinfo{year}{1993}).

\bibitem[{\citenamefont{Muckenheim et~al.}(1986)\citenamefont{Muckenheim,
  Ludwig, Dewdney, Holland, Kyprianidis, Vigier, Cufaro~Petroni, Bartlett, and
  Jaynes}}]{Muckenheim.et.al86}
\bibinfo{author}{\bibfnamefont{W.}~\bibnamefont{Muckenheim}},
  \bibinfo{author}{\bibfnamefont{G.}~\bibnamefont{Ludwig}},
  \bibinfo{author}{\bibfnamefont{C.}~\bibnamefont{Dewdney}},
  \bibinfo{author}{\bibfnamefont{P.~R.} \bibnamefont{Holland}},
  \bibinfo{author}{\bibfnamefont{A.}~\bibnamefont{Kyprianidis}},
  \bibinfo{author}{\bibfnamefont{J.~P.} \bibnamefont{Vigier}},
  \bibinfo{author}{\bibfnamefont{N.}~\bibnamefont{Cufaro~Petroni}},
  \bibinfo{author}{\bibfnamefont{M.~S.} \bibnamefont{Bartlett}},
  \bibnamefont{and} \bibinfo{author}{\bibfnamefont{E.~T.}
  \bibnamefont{Jaynes}}, \bibinfo{journal}{Phys. Rep.}
  \textbf{\bibinfo{volume}{133}}, \bibinfo{pages}{337} (\bibinfo{year}{1986}).

\end{thebibliography}

\end{document}